\documentclass[reprint,aps,pre,superscriptaddress]{revtex4-1}
\usepackage{graphicx}
\usepackage{dcolumn}
\usepackage{bm}
\usepackage{amsmath}
\usepackage{soul}
\usepackage{xcolor}

\usepackage{hyperref}
\hypersetup{
    colorlinks=true,
    linkcolor=black,
    filecolor=black,      
    urlcolor=blue,
    citecolor=black
}

\begin{document}

\title{Bifurcation structure of a swept source laser}

\author{A. V. Kovalev}
\email{avkovalev@niuitmo.ru}
\affiliation{ITMO University, Saint Petersburg, Russia }

\author{P. S. Dmitriev}
\affiliation{ITMO University, Saint Petersburg, Russia }

\author{A. G. Vladimirov}
\affiliation{Weierstrass Institute for Applied Analysis and Stochastics, Mohrenstrasse
39, D-10117 Berlin, Germany}

\author{A. Pimenov}
\affiliation{Weierstrass Institute for Applied Analysis and Stochastics, Mohrenstrasse
39, D-10117 Berlin, Germany}

\author{G. Huyet}
\affiliation{Universit\'e C\^ote d'Azur, CNRS, INPHYNI, France}

\author{E. A. Viktorov}
\affiliation{ITMO University, Saint Petersburg, Russia }

\date{\today }

\begin{abstract}
We numerically analyze a delay differential equation model of a short-cavity
semiconductor laser with an intracavity frequency swept filter and reveal a
complex bifurcation structure responsible for the asymmetry of the output
characteristics of this laser. We show that depending on the direction of
the frequency sweep of a narrowband filter, there exist two bursting cycles
determined by different parts of a continuous-wave solutions branch.

A. V. Kovalev, P. S. Dmitriev, A. G. Vladimirov, A. Pimenov, G. Huyet, and E. A. Viktorov, \href{https://link.aps.org/doi/10.1103/PhysRevE.101.012212}{Phys. Rev. E \textbf{101}, 012212 }(\href{https://link.aps.org/doi/10.1103/PhysRevE.101.012212}{2020}).
\copyright{ 2020 American Physical Society}
\end{abstract}

\maketitle

\section{Introduction}
Optical Coherence Tomography (OCT) has enabled the fast and reliable visualization of various tissues for medical assessment \cite{Drexler2008}. Swept-Source OCT is a technology that relies on coherent lasers that can scan hundreds of nanometers in a few microseconds to enable real time videos and, as a result, has found a wide range of medical applications in areas such as ophthalmology or cardiology \cite{Chinn1997}. To obtain such performance, researchers have
developed novel frequency swept light sources, such as Fourier Domain Mode-Locked Lasers (FDML) \cite%
{Huber}, short external cavity lasers \cite{Axsun1,Axsun2,Johnson2017,Johnson2018},  Vertical
Cavity Surface Emitting Lasers (VCSELs) with micro-electromechanical system (MEMS) driven filters \cite{Fujimoto, Butler2017,Moon2017,Butler2019}, multi-section
semiconductor lasers  \cite{Insight}, and photonic integrated circuit devices \cite{Pajkovic2019}. The underlying operation principle of
these devices relies on laser cavities incorporating a broad band gain
medium and a fast tuning mechanism. Semiconductor quantum well active medium can be
engineered to deliver broadband gain amplification, however, the development of fast
tuning mechanism is a challenge as it may degrade the laser emission. FDML
lasers have a kilometer long ring cavity containing an intracavity filter
that is driven in resonance with the round trip time. At the other extreme,
VCSELs have a cavity length of a single optical wavelength and their
tunability is achieved by a slight modification of the cavity length.

Nonlinear dynamical regimes in FDML devices can be theoretically modeled by
partial differential equations governing the spatio-temporal evolution of
the complex envelope of the electric field \cite{Huber09,Avrutin2019}. Another powerful
method to describe these lasers is based on the use of delay differential
equations (DDEs) \cite{Slepneva13,DDE}. In particular, the experimentally observed asymmetry
in the output dynamics between the filter sweeping from shorter to longer
wavelengths and the filter sweeping from longer to shorter wavelengths has
been successfully explained using the DDE FDML model \cite{Slepneva13}. It
was shown that instabilities observed in FDML lasers can be related to
short- and long-wavelength modulation instabilities commonly found in
nonlinear spatially-distributed systems. The same model was able to describe
the appearance of the so-called \textquotedblleft sliding frequency
mode-locking\textquotedblright\ in short cavity frequency swept lasers \cite%
{Slepneva14}. Shorter cavity length devices are appealing as comparably inexpensive and compact swept OCT sources and have recently attracted significant attention \cite{Johnson2018,Butler2019,Insight,Pajkovic2019}. These lasers, however, demonstrate wide range of dynamical regimes during the filter sweeping \cite{Slepneva14} detrimental to the performance of OCT sources, which were observed only in numerical simulations. Therefore, further analysis and understanding of the dynamical properties of such devices is important for the improvement of their characteristics necessary for the future applications.

Unlike Ref.~\cite{Slepneva13}, where the asymmetry of the FDML laser was
studied in the long cavity limit, in this Letter we consider the case when the cavity
length is relatively small and the free spectral range is larger than the
bandwidth of the tunable filter. We show that in this case the
experimentally observed asymmetry of the laser output with respect to sweep
direction is related to the presence of a fold and Andronov-Hopf
bifurcations of a very asymmetric branch of continuous wave (CW) regimes.
Furthermore, we present a detailed bifurcation analysis of the model
equations, discuss coexisting dynamical regimes such as longitudinal mode
hopping, quasiperiodic pulsations and chaos, and compare the results with
those obtained earlier \cite{Slepneva13} for a long cavity laser.

\section{The model}
We consider a DDE model \cite{Slepneva13} for the normalized complex
amplitude of the electrical field ${\tilde E}$ and the time-dependent
dimensionless cumulative saturable gain $G$:

\begin{eqnarray}
\gamma ^{-1}\frac{d{\tilde{E}}}{dt} &+&\left( 1+i\Delta \right) {\tilde{E}}=%
\sqrt{\kappa }e^{\frac{1-i\alpha }{2}G}{\tilde{E}}\left( t-1\right) ,
\label{P1A} \\
\eta ^{-1}\frac{dG}{dt} &=&J-G-(e^{G}-1)\left\vert {\tilde{E}}\left(
t-1\right) \right\vert ^{2},  \label{P2A}
\end{eqnarray}%
where $t\equiv t^{\prime}/T$, $t^{\prime}$ is time, and $T$ is equal to the cold
cavity round trip time. The attenuation factor $\kappa $ describes the total
non-resonant linear intensity losses per cavity round trip, $\alpha $ is the
linewidth enhancement factor in the gain, and $\gamma $ is the bandwidth of
the intracavity spectral filtering multiplied by the round trip time $T$. $\gamma<1$ $(\gg1)$ corresponds to the short (long) cavity. $J$ is the pump parameter, and $\eta
=\mathcal{O}(1) $ is the ratio of the cold cavity round trip time and the carrier
density relaxation time. The time-dependent parameter $\Delta =\Delta (t)$
defines the detuning between the central frequency of the narrowband tunable
filter and the reference frequency, which coincides with the frequency of
one of the laser modes. After the coordinate change ${\tilde{E}}%
=Ee^{-i\int_{0}^{t}\Delta (x)dx}$, Eqs.~(\ref{P1A}-\ref{P2A}) are transformed
into 
\begin{eqnarray}
\gamma ^{-1}\frac{dE}{dt} &+&E=\sqrt{\kappa }e^{\frac{1-i\alpha }{2}G-i\phi
(t)}E\left( t-1\right) ,  \label{P1} \\
\eta ^{-1}\frac{dG}{dt} &=&J-G-(e^{G}-1)\left\vert E\left( t-1\right)
\right\vert ^{2},  \label{P2}
\end{eqnarray}%
where $\phi (t)=-i\int_{t-1}^{t}\Delta (x)dx$. Note that Eqs.~(\ref{P1}--\ref%
{P2}) are invariant with respect to the shifts $\phi \rightarrow \phi +2\pi
n $, where $n=0,\pm 1,\pm 2\dots $ is an integer number. Therefore, all
bifurcation diagrams studied here are  $2\pi$-periodic on $\phi $.

\begin{figure}[tbp]
\includegraphics[width=\linewidth]{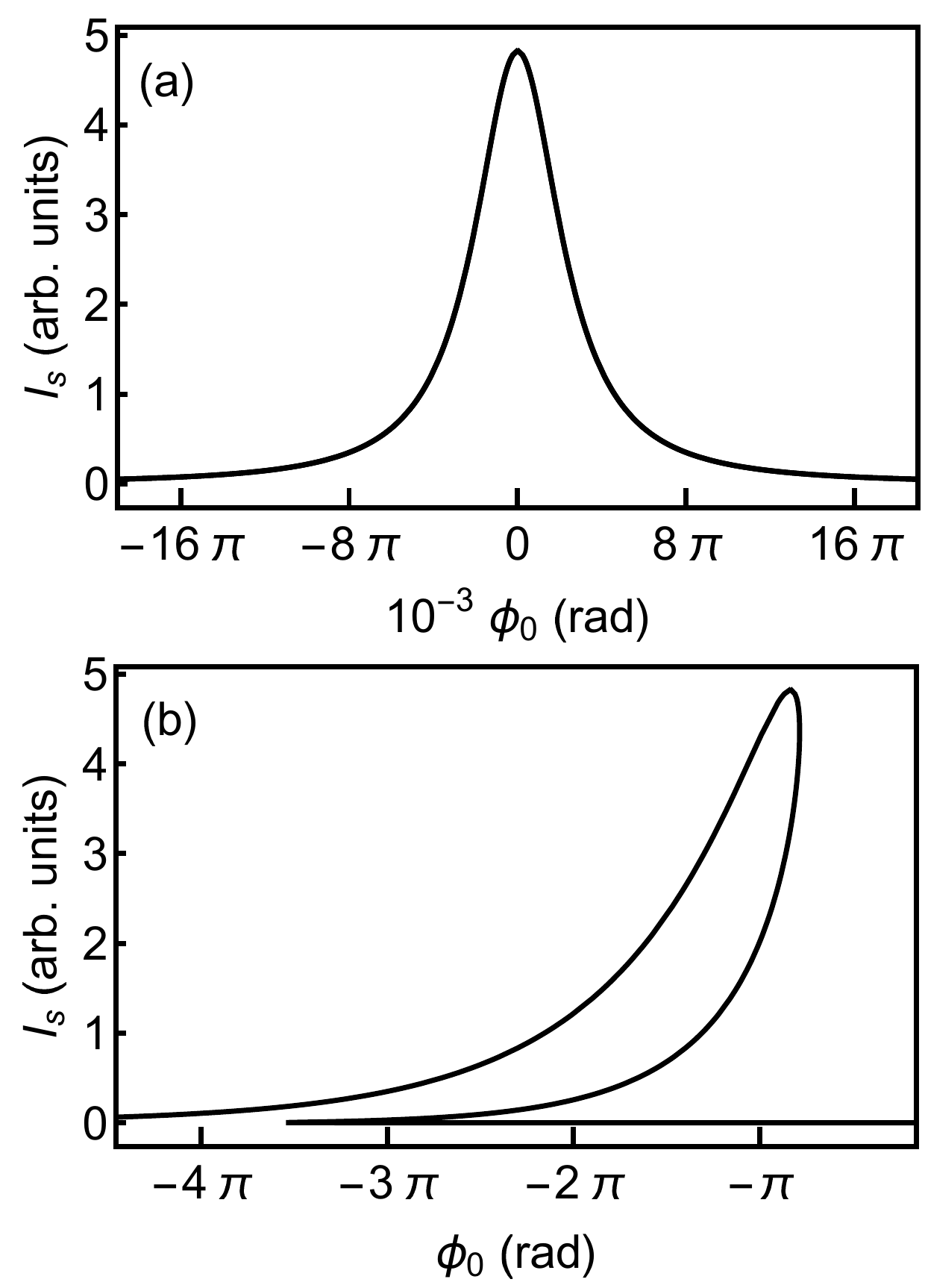}
\caption{Branch of CW solutions in a long cavity ((a), $\protect\gamma =100$%
) and short cavity ((b), $\protect\gamma =0.25$) laser. Other parameter
values are $J=10$, $\protect\kappa =0.35,$ and $\protect\alpha =5.$}
\label{Fig_cw}
\end{figure}

We first consider Eqs.~(\ref{P1}--\ref{P2}) for the static $\phi (t)=\phi
_{0}$ and define the CW cavity mode solution as $E=\sqrt{I_{s}}e^{i\omega t}$ with time independent intensity $I_{s}$, and the constant gain $G=g$.
Different CW solutions correspond to different longitudinal modes of the
laser. The relation between the field intensity $I_s$ and the
value of the saturable gain $g$ is given by 
\begin{equation}
I_{s}=\frac{J-g}{e^{g}-1}  \label{I}.
\end{equation}%
By solving this equation with respect to the gain, $g=g(I_{s})$, we obtain
two values of the modal frequency corresponding to a given value of the
intensity $I_{s}$: 
\begin{equation}
\omega =\pm \gamma \left[ \kappa e^{g(I_{s})}-1\right] .  \label{omega}
\end{equation}%
Finally, substituting Eq.~(\ref{omega}) into the transcendental equation 
\begin{equation}
\phi _{0}=-\omega-\frac{\alpha g(I_{s})}{2}-\arctan \left( \frac{\omega }{%
\gamma }\right) +2\pi n,  \label{cw}
\end{equation}%
%
with $n=0,\pm 1,\pm 2\dots $, we get an implicit equation relating the
intensity $I_{s}$ and the parameter $\phi _{0}$. The branch of CW solutions
defined by Eqs.~(\ref{I}--\ref{cw}) with $n=0$ is shown in Fig.~\ref{Fig_cw}
for the case of long cavity (a) and short cavity (b)
laser. All other CW branches can be obtained by a shift $\phi
_{0}\rightarrow \phi _{0}+2\pi n$ with integer $n$. It is seen that in a
long cavity laser studied in \cite{Slepneva13}, the CW branch is almost
symmetric with respect to the reflection $\phi _{0}\rightarrow -\phi _{0}$.

In a short cavity laser, the CW branch can be very asymmetric with a foldover, which is
generally characteristic for nonlinear resonators \cite{Otsuka1983,Coen}. The fold bifurcation points  in the Fig.~\ref{Fig_cw}(b), corresponding to the extrema of the  function $\phi_0(\omega)$ defined by (\ref{cw}), can be found by solving $d\phi_0/d\omega = 0$  and read:
\begin{equation}
 \omega_{LP}=(-\alpha \pm \sqrt{\alpha ^2-4 \gamma (\gamma +1)})/2. 
 \label{eq_fold}
\end{equation}
Inequality $\alpha ^{2}>4\gamma \left( 1+\gamma \right)$ defines the condition for appearance of the foldover.
One of the two fold points defined by (\ref{eq_fold}) corresponds to the small
intensity and another to the large intensity, as can be seen in the Fig.~\ref{Fig_cw}(b). The latter fold bifurcation is responsible for
the stability loss of a CW regime in a laser with adiabatically slowly
increasing $\phi _{0}$.

\begin{figure}[tbp]
\includegraphics[width=\linewidth]{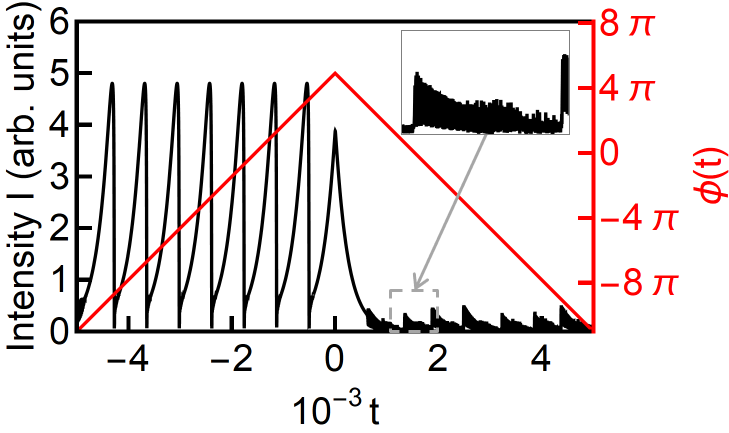}
\caption{Numerical simulation of the model equations (3-4) displaying mode-hopping events in the positive sweep direction. The frequency sweeping in the negative direction exhibits chaotic dynamics. The zero point on the x-axis is the turning point of the sweep and the sweeping function $\phi(t)$ is shown (in red) above the intensity. The parameters are: $\eta=1$, $\gamma=0.25$, $\varepsilon=0.01$. The other parameter
values are the same as in Fig.~\ref{Fig_cw}.}
\label{Fig_tt}
\end{figure}

\begin{figure}[tbp]
\includegraphics[width=\linewidth]{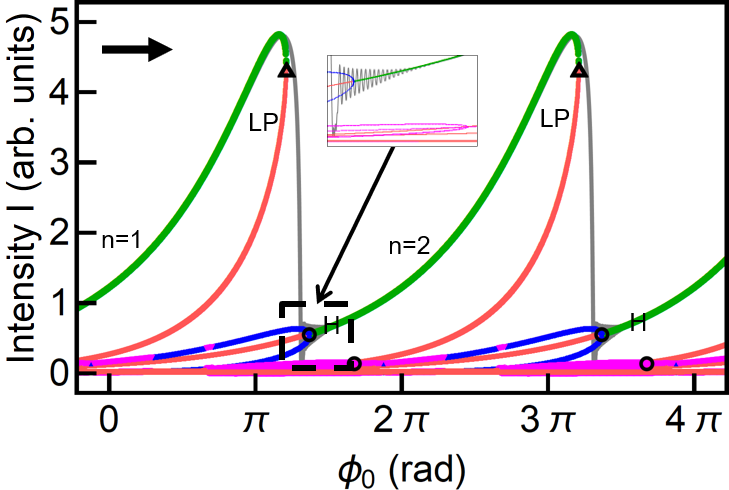}
\caption{The power-dropout/power-recovery large amplitude cycle (dark grey)
in the plane $(I,\protect\phi _{0})$ is shown together with the
bifurcation diagram of the cavity modes $n=1$ and $n=2$ in the interval $0<%
\protect\phi _{0}<4\protect\pi .$ Green (red) lines correspond to the stable
(unstable) steady state solutions. Blue (magenta) lines correspond to the
stable (unstable) periodic solutions. Circles and triangles mark an
Andronov-Hopf bifurcation point and a fold bifurcation, respectively. The
figure shows that the power-dropout/power-recovery cycle follows a stable
branch of periodic solutions until it reaches a supercritical Andronov-Hopf
bifurcation point $H$, than follows the stable steady state branch until it
reaches a fold bifurcation point $LP$. The black arrow indicates the
direction of sweep. The values of the fixed parameters are the same as in
Fig.~\ref{Fig_tt}.}
\label{Fig3}
\end{figure}

\section{Sweeping dynamics}
Let us now explore the effect of a slowly varying $\phi (t)=\pm \varepsilon
t$, $\varepsilon \ll 1$ that corresponds to the frequency sweep in opposite
directions with a sweeping rate which is much slower than one wavelength per round trip. Time trace in Fig.~\ref{Fig_tt} results from direct numerical integration of Eqs. (\ref{P1}--\ref{P2}), and demonstrates well known asymmetry of the dynamical response to the frequency sweep. The bifurcation diagram of the steady and periodic solutions in
Fig.~\ref{Fig3} has been computed using a numerical continuation technique \cite%
{ddebitool} and displays the cavity mode branches for $n=1$ and $n=2$ in the
range $0<\phi _{0}<4\pi .$ Because of the periodicity in $\phi _{0}$, the
cavity mode branch for $n=2$ is the same as the one for $n=1$ but shifted by
$2\pi $\ along the $\phi _{0}$-axis. The low amplitude tail of the branch for $n=2$
overlaps with the large amplitude part of the branch for $n=1.$ This overlap is
important for understanding of two types of the bursting dynamics which
appear with frequency sweeping in opposite directions. Each branch contains
two important bifurcations marked in Fig.~\ref{Fig3} as $H$ and $LP$, and a stable
steady state laser operation is only possible in the interval between these
points. $\ LP$ corresponds to a fold bifurcation from a cavity mode that is
responsible for the mode hopping sequence as we progressively increase $\phi
_{0}$. The mode hopping sequence forms large amplitude bursts which are
similar to neuromorphic design of square-wave bursting oscillations \cite%
{eugene}.

\begin{figure}[tb]
\includegraphics[width=\linewidth]{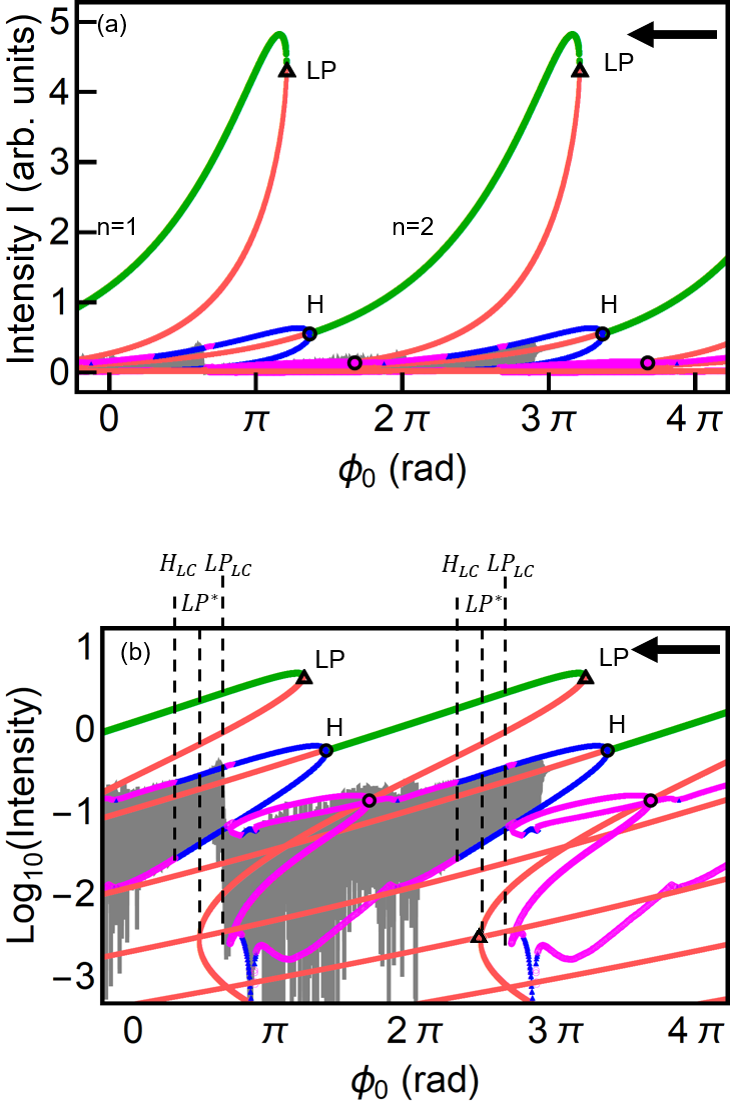}
\caption{Dynamical evolution of the intensity with sweeping frequency in the
positive direction (dark gray) in the plane $(I,\protect\phi _{0})$ is
shown together with the bifurcation diagram in linear (a) and logarithmic scales
(b). The figure shows that the branch of periodic solutions (dark grey) is
emerging from supercritical Andronov-Hopf bifurcation point $H$, follows a stable
branch of periodic solutions, undergoes a secondary Hopf bifurcation $H_{LC}$, and develops into chaos with various stability
changes until it reaches a limit-point of limit-cycles $LP_{LC}$ from where
it jumps back to the vicinity of the Andronov-Hopf bifurcation. The black arrow
indicates the direction of sweep. $LP^{*}$ is the CW solution fold bifurcation point at low intensity value. The coloring, the marks and the fixed
parameters are the same as in Fig.~\ref{Fig3}.}
\label{Fig4}
\end{figure}

Formation of the large amplitude burst is detailed in Fig.~\ref{Fig3} where the
bifurcation diagram of the steady state and periodic solutions is shown
together with the long time solution of Eqs. (\ref{P1}-\ref{P2}) (in dark
gray) for the positive frequency sweep direction relative to the filter profile $\phi (t)=\varepsilon t,\varepsilon =0.01$. The single mode
steady state changes stability at the point $H$ with the increase of $\phi
_{0}$, and the branch of stable periodic solutions emerges from the
supercritical Andronov-Hopf bifurcation point at the relaxation oscillation
frequency. $LP$ marks a limit point of steady states at which the power
dropout happens. The laser follows the steady state branch $n=1$ as $\phi
_{0}$ increases until it passes $LP$, and then drops down to sustained
oscillations of the lower branch of periodic solutions at $n=2$, and returns
to the steady state branch passing the Andronov-Hopf bifurcation $H.$ As is
visible in Fig.~\ref{Fig3}, the Andronov-Hopf bifurcation transition to steady state
can be delayed in the absence of noise \cite{baer}. $\ $

\begin{figure}[tbp]
\includegraphics[width=\linewidth]{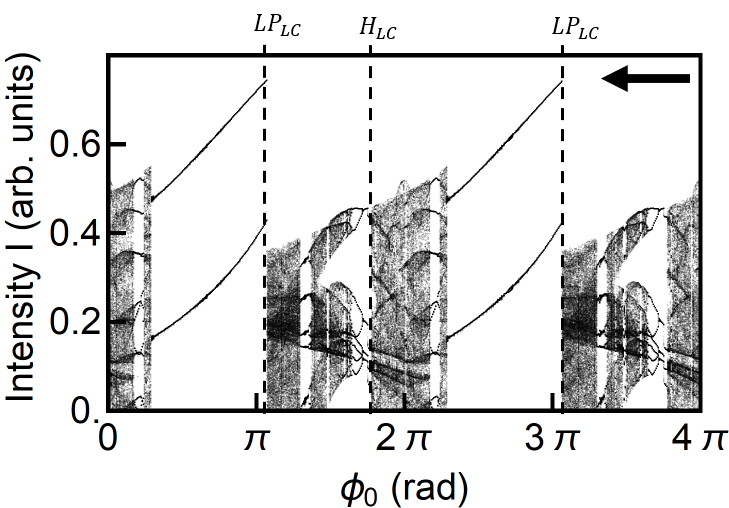}
\caption{Time-average extrema of $I(t)$ obtained numerically for the
negative direction of the frequency sweep for low amplitude bursting. The
black arrow indicates the direction of sweep. The fixed parameters and the
marks are the same as in Fig.~\ref{Fig4}.}
\label{Fig5}
\end{figure}

Let us now follow a low amplitude bursting cycle which appears at the branch 
$n=2$ after a supercritical Andronov-Hopf bifurcation $H$ for the
negative sweep direction $\phi (t)=-\varepsilon t$, $\varepsilon =0.01$. It is
shown in dark grey in Fig.~\ref{Fig4}. After the transition to the stable periodic
oscillations the laser follows the branch \ $n=2$ of limit-cycle
oscillations as $\phi _{0}$ decreases until it reaches $LP_{LC}$.\ The laser
then jumps up to the upper branch $n=1$, starting a new bursting cycle. The
jump up may happen slightly before $LP_{LC}.$ The folding point of the
Andronov-Hopf bifurcation branch, which we denote by $LP_{LC}$ in Fig.~\ref{Fig4}(b), is
important for the formation of the low amplitude bursting. This point
corresponds to a saddle-node bifurcation of limit cycles below which neither
stable nor unstable periodic oscillations are possible. Different dynamics
between $H$ and $LP_{LC}$ can be seen in Fig.~\ref{Fig5} which shows the extrema of
the oscillations as we progressively decrease $\phi _{0}$ from $H$. After a
secondary Hopf bifurcation $H_{LC}$, quasiperiodicity and a weak chaos, the
laser jumps up to the higher branch. The response of the laser to the slowly sweeping narrow band filtering thus takes the form of low amplitude bursts of spiking.

\section{Conclusion}
In this paper, we have considered a delay differential equation model for a laser with an intracavity swept filter, and theoretically analyzed the bifurcation structure of a short cavity swept source. Unlike the long cavity devices, the continuous wave solution of the model equations is strongly asymmetric with a foldover similar to nonlinear resonance curve with hysteresis \cite{Otsuka1983,Coen}. The foldover allows coexistence of single mode branches what changes the character of the mode hopping compared to long cavity devices. Additionally, the foldover defines two bursting phenomena which form sufficiently different laser outputs depending on the sweep direction. Such a behavior is similar to that observed in other swept sources; for this reason the increasing wavelength sweep
will lead to more coherent output but with mode hops, while the
decreasing wavelength sweep will lead to a continuous sweep
with a lower coherence length as for other swept sources  \cite{Slepneva13,Slepneva14}.

\begin{acknowledgments}
A.V.K., P.S.D., and E.A.V. acknowledge the Government of the Russian Federation (Grant 08-08). A.V.K., P.S.D., G.H., and E.A.V. acknowledge the support of H2020-MSCA-RISE-2018 HALT. A.V.K. and E.A.V. work was supported by the Ministry of Education and Science of the Russian Federation (Passport No. 2019-1442). A.P. and A.G.V. acknowledge the support by the subproject B5 of the DFG Collaborative Research Center SFB 787. A.P. is funded by the DFG under Germanys Excellence Strategy—The Berlin Mathematics Research Center MATH+ (EXC-2046/1, project ID: 390685689).
\end{acknowledgments}

\bibliography{BifSwepSource16}

\begin{thebibliography}{23}%
\makeatletter
\providecommand \@ifxundefined [1]{%
 \@ifx{#1\undefined}
}%
\providecommand \@ifnum [1]{%
 \ifnum #1\expandafter \@firstoftwo
 \else \expandafter \@secondoftwo
 \fi
}%
\providecommand \@ifx [1]{%
 \ifx #1\expandafter \@firstoftwo
 \else \expandafter \@secondoftwo
 \fi
}%
\providecommand \natexlab [1]{#1}%
\providecommand \enquote  [1]{``#1''}%
\providecommand \bibnamefont  [1]{#1}%
\providecommand \bibfnamefont [1]{#1}%
\providecommand \citenamefont [1]{#1}%
\providecommand \href@noop [0]{\@secondoftwo}%
\providecommand \href [0]{\begingroup \@sanitize@url \@href}%
\providecommand \@href[1]{\@@startlink{#1}\@@href}%
\providecommand \@@href[1]{\endgroup#1\@@endlink}%
\providecommand \@sanitize@url [0]{\catcode `\\12\catcode `\$12\catcode
  `\&12\catcode `\#12\catcode `\^12\catcode `\_12\catcode `\%12\relax}%
\providecommand \@@startlink[1]{}%
\providecommand \@@endlink[0]{}%
\providecommand \url  [0]{\begingroup\@sanitize@url \@url }%
\providecommand \@url [1]{\endgroup\@href {#1}{\urlprefix }}%
\providecommand \urlprefix  [0]{URL }%
\providecommand \Eprint [0]{\href }%
\providecommand \doibase [0]{http://dx.doi.org/}%
\providecommand \selectlanguage [0]{\@gobble}%
\providecommand \bibinfo  [0]{\@secondoftwo}%
\providecommand \bibfield  [0]{\@secondoftwo}%
\providecommand \translation [1]{[#1]}%
\providecommand \BibitemOpen [0]{}%
\providecommand \bibitemStop [0]{}%
\providecommand \bibitemNoStop [0]{.\EOS\space}%
\providecommand \EOS [0]{\spacefactor3000\relax}%
\providecommand \BibitemShut  [1]{\csname bibitem#1\endcsname}%
\let\auto@bib@innerbib\@empty
\bibitem [{\citenamefont {Drexler}\ and\ \citenamefont
  {Fujimoto}(2008)}]{Drexler2008}%
  \BibitemOpen
  \bibinfo {editor} {\bibfnamefont {W.}~\bibnamefont {Drexler}}\ and\ \bibinfo
  {editor} {\bibfnamefont {J.~G.}\ \bibnamefont {Fujimoto}},\ eds.,\ \href
  {\doibase 10.1007/978-3-540-77550-8} {\emph {\bibinfo {title} {Optical
  Coherence Tomography}}}\ (\bibinfo  {publisher} {Springer Berlin
  Heidelberg},\ \bibinfo {year} {2008})\BibitemShut {NoStop}%
\bibitem [{\citenamefont {Chinn}\ \emph {et~al.}(1997)\citenamefont {Chinn},
  \citenamefont {Swanson},\ and\ \citenamefont {Fujimoto}}]{Chinn1997}%
  \BibitemOpen
  \bibfield  {author} {\bibinfo {author} {\bibfnamefont {S.~R.}\ \bibnamefont
  {Chinn}}, \bibinfo {author} {\bibfnamefont {E.~A.}\ \bibnamefont {Swanson}},
  \ and\ \bibinfo {author} {\bibfnamefont {J.~G.}\ \bibnamefont {Fujimoto}},\
  }\href {\doibase 10.1364/ol.22.000340} {\bibfield  {journal} {\bibinfo
  {journal} {Optics Letters}\ }\textbf {\bibinfo {volume} {22}},\ \bibinfo
  {pages} {340} (\bibinfo {year} {1997})}\BibitemShut {NoStop}%
\bibitem [{\citenamefont {Adler}\ \emph {et~al.}(2011)\citenamefont {Adler},
  \citenamefont {Wieser}, \citenamefont {Trepanier}, \citenamefont {Schmitt},\
  and\ \citenamefont {Huber}}]{Huber}%
  \BibitemOpen
  \bibfield  {author} {\bibinfo {author} {\bibfnamefont {D.~C.}\ \bibnamefont
  {Adler}}, \bibinfo {author} {\bibfnamefont {W.}~\bibnamefont {Wieser}},
  \bibinfo {author} {\bibfnamefont {F.}~\bibnamefont {Trepanier}}, \bibinfo
  {author} {\bibfnamefont {J.~M.}\ \bibnamefont {Schmitt}}, \ and\ \bibinfo
  {author} {\bibfnamefont {R.~A.}\ \bibnamefont {Huber}},\ }\href {\doibase
  10.1364/oe.19.020930} {\bibfield  {journal} {\bibinfo  {journal} {Optics
  Express}\ }\textbf {\bibinfo {volume} {19}},\ \bibinfo {pages} {20930}
  (\bibinfo {year} {2011})}\BibitemShut {NoStop}%
\bibitem [{\citenamefont {Atia}\ \emph {et~al.}(2009)\citenamefont {Atia},
  \citenamefont {Kuznetsov},\ and\ \citenamefont {Flanders}}]{Axsun1}%
  \BibitemOpen
  \bibfield  {author} {\bibinfo {author} {\bibfnamefont {W.}~\bibnamefont
  {Atia}}, \bibinfo {author} {\bibfnamefont {M.}~\bibnamefont {Kuznetsov}}, \
  and\ \bibinfo {author} {\bibfnamefont {D.}~\bibnamefont {Flanders}},\
  }\href@noop {} {\enquote {\bibinfo {title} {Linearized swept laser source for
  optical coherence analysis system},}\ } (\bibinfo {year} {2009}),\ \bibinfo
  {note} {uS Patent App. 12/027,710.}\BibitemShut {Stop}%
\bibitem [{\citenamefont {Johnson}\ and\ \citenamefont
  {Flanders}(2013)}]{Axsun2}%
  \BibitemOpen
  \bibfield  {author} {\bibinfo {author} {\bibfnamefont {B.}~\bibnamefont
  {Johnson}}\ and\ \bibinfo {author} {\bibfnamefont {D.}~\bibnamefont
  {Flanders}},\ }\href@noop {} {\enquote {\bibinfo {title} {Laser swept source
  with controlled mode locking for oct medical imaging},}\ } (\bibinfo {year}
  {2013}),\ \bibinfo {note} {eP Patent App. EP20,110,808,812}\BibitemShut
  {NoStop}%
\bibitem [{\citenamefont {Johnson}\ \emph {et~al.}(2017)\citenamefont
  {Johnson}, \citenamefont {Atia}, \citenamefont {Kuznetsov}, \citenamefont
  {Goldberg}, \citenamefont {Whitney},\ and\ \citenamefont
  {Flanders}}]{Johnson2017}%
  \BibitemOpen
  \bibfield  {author} {\bibinfo {author} {\bibfnamefont {B.}~\bibnamefont
  {Johnson}}, \bibinfo {author} {\bibfnamefont {W.}~\bibnamefont {Atia}},
  \bibinfo {author} {\bibfnamefont {M.}~\bibnamefont {Kuznetsov}}, \bibinfo
  {author} {\bibfnamefont {B.~D.}\ \bibnamefont {Goldberg}}, \bibinfo {author}
  {\bibfnamefont {P.}~\bibnamefont {Whitney}}, \ and\ \bibinfo {author}
  {\bibfnamefont {D.~C.}\ \bibnamefont {Flanders}},\ }\href {\doibase
  10.1364/boe.8.001045} {\bibfield  {journal} {\bibinfo  {journal} {Biomedical
  Optics Express}\ }\textbf {\bibinfo {volume} {8}},\ \bibinfo {pages} {1045}
  (\bibinfo {year} {2017})}\BibitemShut {NoStop}%
\bibitem [{\citenamefont {Johnson}\ \emph {et~al.}(2018)\citenamefont
  {Johnson}, \citenamefont {Atia}, \citenamefont {Kuznetsov}, \citenamefont
  {Goldberg}, \citenamefont {Whitney},\ and\ \citenamefont
  {Flanders}}]{Johnson2018}%
  \BibitemOpen
  \bibfield  {author} {\bibinfo {author} {\bibfnamefont {B.}~\bibnamefont
  {Johnson}}, \bibinfo {author} {\bibfnamefont {W.}~\bibnamefont {Atia}},
  \bibinfo {author} {\bibfnamefont {M.}~\bibnamefont {Kuznetsov}}, \bibinfo
  {author} {\bibfnamefont {B.~D.}\ \bibnamefont {Goldberg}}, \bibinfo {author}
  {\bibfnamefont {P.}~\bibnamefont {Whitney}}, \ and\ \bibinfo {author}
  {\bibfnamefont {D.~C.}\ \bibnamefont {Flanders}},\ }\href {\doibase
  10.1364/oe.26.034909} {\bibfield  {journal} {\bibinfo  {journal} {Optics
  Express}\ }\textbf {\bibinfo {volume} {26}},\ \bibinfo {pages} {34909}
  (\bibinfo {year} {2018})}\BibitemShut {NoStop}%
\bibitem [{\citenamefont {Jayaraman}\ \emph {et~al.}(2012)\citenamefont
  {Jayaraman}, \citenamefont {Cole}, \citenamefont {Robertson}, \citenamefont
  {Uddin},\ and\ \citenamefont {Cable}}]{Fujimoto}%
  \BibitemOpen
  \bibfield  {author} {\bibinfo {author} {\bibfnamefont {V.}~\bibnamefont
  {Jayaraman}}, \bibinfo {author} {\bibfnamefont {G.}~\bibnamefont {Cole}},
  \bibinfo {author} {\bibfnamefont {M.}~\bibnamefont {Robertson}}, \bibinfo
  {author} {\bibfnamefont {A.}~\bibnamefont {Uddin}}, \ and\ \bibinfo {author}
  {\bibfnamefont {A.}~\bibnamefont {Cable}},\ }\href {\doibase
  10.1049/el.2012.1552} {\bibfield  {journal} {\bibinfo  {journal} {Electronics
  Letters}\ }\textbf {\bibinfo {volume} {48}},\ \bibinfo {pages} {867}
  (\bibinfo {year} {2012})}\BibitemShut {NoStop}%
\bibitem [{\citenamefont {Butler}\ \emph {et~al.}(2017)\citenamefont {Butler},
  \citenamefont {Slepneva}, \citenamefont {McNamara}, \citenamefont {Neuhaus},
  \citenamefont {Goulding}, \citenamefont {Leahy},\ and\ \citenamefont
  {Huyet}}]{Butler2017}%
  \BibitemOpen
  \bibfield  {author} {\bibinfo {author} {\bibfnamefont {T.~P.}\ \bibnamefont
  {Butler}}, \bibinfo {author} {\bibfnamefont {S.}~\bibnamefont {Slepneva}},
  \bibinfo {author} {\bibfnamefont {P.~M.}\ \bibnamefont {McNamara}}, \bibinfo
  {author} {\bibfnamefont {K.}~\bibnamefont {Neuhaus}}, \bibinfo {author}
  {\bibfnamefont {D.}~\bibnamefont {Goulding}}, \bibinfo {author}
  {\bibfnamefont {M.}~\bibnamefont {Leahy}}, \ and\ \bibinfo {author}
  {\bibfnamefont {G.}~\bibnamefont {Huyet}},\ }\href {\doibase
  10.1109/jphot.2017.2752644} {\bibfield  {journal} {\bibinfo  {journal}
  {{IEEE} Photonics Journal}\ }\textbf {\bibinfo {volume} {9}},\ \bibinfo
  {pages} {1} (\bibinfo {year} {2017})}\BibitemShut {NoStop}%
\bibitem [{\citenamefont {Moon}\ and\ \citenamefont {Choi}(2017)}]{Moon2017}%
  \BibitemOpen
  \bibfield  {author} {\bibinfo {author} {\bibfnamefont {S.}~\bibnamefont
  {Moon}}\ and\ \bibinfo {author} {\bibfnamefont {E.~S.}\ \bibnamefont
  {Choi}},\ }\href {\doibase 10.1364/boe.8.001110} {\bibfield  {journal}
  {\bibinfo  {journal} {Biomedical Optics Express}\ }\textbf {\bibinfo {volume}
  {8}},\ \bibinfo {pages} {1110} (\bibinfo {year} {2017})}\BibitemShut
  {NoStop}%
\bibitem [{\citenamefont {Butler}\ \emph {et~al.}(2019)\citenamefont {Butler},
  \citenamefont {Goulding}, \citenamefont {Slepneva}, \citenamefont
  {O'Shaughnessy}, \citenamefont {Hegarty}, \citenamefont {Huyet},\ and\
  \citenamefont {Kelleher}}]{Butler2019}%
  \BibitemOpen
  \bibfield  {author} {\bibinfo {author} {\bibfnamefont {T.~P.}\ \bibnamefont
  {Butler}}, \bibinfo {author} {\bibfnamefont {D.}~\bibnamefont {Goulding}},
  \bibinfo {author} {\bibfnamefont {S.}~\bibnamefont {Slepneva}}, \bibinfo
  {author} {\bibfnamefont {B.}~\bibnamefont {O'Shaughnessy}}, \bibinfo {author}
  {\bibfnamefont {S.~P.}\ \bibnamefont {Hegarty}}, \bibinfo {author}
  {\bibfnamefont {G.}~\bibnamefont {Huyet}}, \ and\ \bibinfo {author}
  {\bibfnamefont {B.}~\bibnamefont {Kelleher}},\ }\href {\doibase
  10.1364/oe.27.007307} {\bibfield  {journal} {\bibinfo  {journal} {Optics
  Express}\ }\textbf {\bibinfo {volume} {27}},\ \bibinfo {pages} {7307}
  (\bibinfo {year} {2019})}\BibitemShut {NoStop}%
\bibitem [{\citenamefont {Bonesi}\ \emph {et~al.}(2014)\citenamefont {Bonesi},
  \citenamefont {Minneman}, \citenamefont {Ensher}, \citenamefont {Zabihian},
  \citenamefont {Sattmann}, \citenamefont {Boschert}, \citenamefont {Hoover},
  \citenamefont {Leitgeb}, \citenamefont {Crawford},\ and\ \citenamefont
  {Drexler}}]{Insight}%
  \BibitemOpen
  \bibfield  {author} {\bibinfo {author} {\bibfnamefont {M.}~\bibnamefont
  {Bonesi}}, \bibinfo {author} {\bibfnamefont {M.~P.}\ \bibnamefont
  {Minneman}}, \bibinfo {author} {\bibfnamefont {J.}~\bibnamefont {Ensher}},
  \bibinfo {author} {\bibfnamefont {B.}~\bibnamefont {Zabihian}}, \bibinfo
  {author} {\bibfnamefont {H.}~\bibnamefont {Sattmann}}, \bibinfo {author}
  {\bibfnamefont {P.}~\bibnamefont {Boschert}}, \bibinfo {author}
  {\bibfnamefont {E.}~\bibnamefont {Hoover}}, \bibinfo {author} {\bibfnamefont
  {R.~A.}\ \bibnamefont {Leitgeb}}, \bibinfo {author} {\bibfnamefont
  {M.}~\bibnamefont {Crawford}}, \ and\ \bibinfo {author} {\bibfnamefont
  {W.}~\bibnamefont {Drexler}},\ }\href {\doibase 10.1364/oe.22.002632}
  {\bibfield  {journal} {\bibinfo  {journal} {Optics Express}\ }\textbf
  {\bibinfo {volume} {22}},\ \bibinfo {pages} {2632} (\bibinfo {year}
  {2014})}\BibitemShut {NoStop}%
\bibitem [{\citenamefont {Pajkovi{\'{c}}}\ \emph {et~al.}(2019)\citenamefont
  {Pajkovi{\'{c}}}, \citenamefont {Tian}, \citenamefont {Latkowski},
  \citenamefont {Williams},\ and\ \citenamefont {Bente}}]{Pajkovic2019}%
  \BibitemOpen
  \bibfield  {author} {\bibinfo {author} {\bibfnamefont {R.}~\bibnamefont
  {Pajkovi{\'{c}}}}, \bibinfo {author} {\bibfnamefont {Y.}~\bibnamefont
  {Tian}}, \bibinfo {author} {\bibfnamefont {S.}~\bibnamefont {Latkowski}},
  \bibinfo {author} {\bibfnamefont {K.~A.}\ \bibnamefont {Williams}}, \ and\
  \bibinfo {author} {\bibfnamefont {E.~A.~M.}\ \bibnamefont {Bente}},\ }in\
  \href {\doibase 10.1117/12.2509572} {\emph {\bibinfo {booktitle} {Novel
  In-Plane Semiconductor Lasers {XVIII}}}},\ \bibinfo {editor} {edited by\
  \bibinfo {editor} {\bibfnamefont {A.~A.}\ \bibnamefont {Belyanin}}\ and\
  \bibinfo {editor} {\bibfnamefont {P.~M.}\ \bibnamefont {Smowton}}}\ (\bibinfo
   {publisher} {{SPIE}},\ \bibinfo {year} {2019})\BibitemShut {NoStop}%
\bibitem [{\citenamefont {Jirauschek}\ \emph {et~al.}(2009)\citenamefont
  {Jirauschek}, \citenamefont {Biedermann},\ and\ \citenamefont
  {Huber}}]{Huber09}%
  \BibitemOpen
  \bibfield  {author} {\bibinfo {author} {\bibfnamefont {C.}~\bibnamefont
  {Jirauschek}}, \bibinfo {author} {\bibfnamefont {B.}~\bibnamefont
  {Biedermann}}, \ and\ \bibinfo {author} {\bibfnamefont {R.}~\bibnamefont
  {Huber}},\ }\href {\doibase 10.1364/oe.17.024013} {\bibfield  {journal}
  {\bibinfo  {journal} {Optics Express}\ }\textbf {\bibinfo {volume} {17}},\
  \bibinfo {pages} {24013} (\bibinfo {year} {2009})}\BibitemShut {NoStop}%
\bibitem [{\citenamefont {Avrutin}\ and\ \citenamefont
  {Zhang}(2019)}]{Avrutin2019}%
  \BibitemOpen
  \bibfield  {author} {\bibinfo {author} {\bibfnamefont {E.~A.}\ \bibnamefont
  {Avrutin}}\ and\ \bibinfo {author} {\bibfnamefont {L.}~\bibnamefont
  {Zhang}},\ }\href {\doibase 10.1140/epjb/e2019-90639-3} {\bibfield  {journal}
  {\bibinfo  {journal} {The European Physical Journal B}\ }\textbf {\bibinfo
  {volume} {92}} (\bibinfo {year} {2019}),\
  10.1140/epjb/e2019-90639-3}\BibitemShut {NoStop}%
\bibitem [{\citenamefont {Slepneva}\ \emph {et~al.}(2013)\citenamefont
  {Slepneva}, \citenamefont {Kelleher}, \citenamefont {O'Shaughnessy},
  \citenamefont {Hegarty}, \citenamefont {Vladimirov},\ and\ \citenamefont
  {Huyet}}]{Slepneva13}%
  \BibitemOpen
  \bibfield  {author} {\bibinfo {author} {\bibfnamefont {S.}~\bibnamefont
  {Slepneva}}, \bibinfo {author} {\bibfnamefont {B.}~\bibnamefont {Kelleher}},
  \bibinfo {author} {\bibfnamefont {B.}~\bibnamefont {O'Shaughnessy}}, \bibinfo
  {author} {\bibfnamefont {S.}~\bibnamefont {Hegarty}}, \bibinfo {author}
  {\bibfnamefont {A.}~\bibnamefont {Vladimirov}}, \ and\ \bibinfo {author}
  {\bibfnamefont {G.}~\bibnamefont {Huyet}},\ }\href {\doibase
  10.1364/OE.21.019240} {\bibfield  {journal} {\bibinfo  {journal} {Opt.
  Express}\ }\textbf {\bibinfo {volume} {21}},\ \bibinfo {pages} {19240}
  (\bibinfo {year} {2013})}\BibitemShut {NoStop}%
\bibitem [{\citenamefont {Vladimirov}\ and\ \citenamefont
  {Turaev}(2005)}]{DDE}%
  \BibitemOpen
  \bibfield  {author} {\bibinfo {author} {\bibfnamefont {A.~G.}\ \bibnamefont
  {Vladimirov}}\ and\ \bibinfo {author} {\bibfnamefont {D.}~\bibnamefont
  {Turaev}},\ }\href@noop {} {\bibfield  {journal} {\bibinfo  {journal}
  {Physical Review A}\ }\textbf {\bibinfo {volume} {72}},\ \bibinfo {pages}
  {033808} (\bibinfo {year} {2005})}\BibitemShut {NoStop}%
\bibitem [{\citenamefont {Slepneva}\ \emph {et~al.}(2014)\citenamefont
  {Slepneva}, \citenamefont {O'Shaughnessy}, \citenamefont {Kelleher},
  \citenamefont {Hegarty}, \citenamefont {Vladimirov}, \citenamefont {Lyu},
  \citenamefont {Karnowski}, \citenamefont {Wojtkowski},\ and\ \citenamefont
  {Huyet}}]{Slepneva14}%
  \BibitemOpen
  \bibfield  {author} {\bibinfo {author} {\bibfnamefont {S.}~\bibnamefont
  {Slepneva}}, \bibinfo {author} {\bibfnamefont {B.}~\bibnamefont
  {O'Shaughnessy}}, \bibinfo {author} {\bibfnamefont {B.}~\bibnamefont
  {Kelleher}}, \bibinfo {author} {\bibfnamefont {S.}~\bibnamefont {Hegarty}},
  \bibinfo {author} {\bibfnamefont {A.}~\bibnamefont {Vladimirov}}, \bibinfo
  {author} {\bibfnamefont {H.-C.}\ \bibnamefont {Lyu}}, \bibinfo {author}
  {\bibfnamefont {K.}~\bibnamefont {Karnowski}}, \bibinfo {author}
  {\bibfnamefont {M.}~\bibnamefont {Wojtkowski}}, \ and\ \bibinfo {author}
  {\bibfnamefont {G.}~\bibnamefont {Huyet}},\ }\href {\doibase
  10.1364/OE.22.018177} {\bibfield  {journal} {\bibinfo  {journal} {Opt.
  Express}\ }\textbf {\bibinfo {volume} {22}},\ \bibinfo {pages} {18177}
  (\bibinfo {year} {2014})}\BibitemShut {NoStop}%
\bibitem [{\citenamefont {Otsuka}\ and\ \citenamefont
  {Kobayashi}(1983)}]{Otsuka1983}%
  \BibitemOpen
  \bibfield  {author} {\bibinfo {author} {\bibfnamefont {K.}~\bibnamefont
  {Otsuka}}\ and\ \bibinfo {author} {\bibfnamefont {S.}~\bibnamefont
  {Kobayashi}},\ }\href {\doibase 10.1049/el:19830181} {\bibfield  {journal}
  {\bibinfo  {journal} {Electronics Letters}\ }\textbf {\bibinfo {volume}
  {19}},\ \bibinfo {pages} {262} (\bibinfo {year} {1983})}\BibitemShut
  {NoStop}%
\bibitem [{\citenamefont {Coen}\ and\ \citenamefont {Erkintalo}(2013)}]{Coen}%
  \BibitemOpen
  \bibfield  {author} {\bibinfo {author} {\bibfnamefont {S.}~\bibnamefont
  {Coen}}\ and\ \bibinfo {author} {\bibfnamefont {M.}~\bibnamefont
  {Erkintalo}},\ }\href {\doibase 10.1364/ol.38.001790} {\bibfield  {journal}
  {\bibinfo  {journal} {Optics Letters}\ }\textbf {\bibinfo {volume} {38}},\
  \bibinfo {pages} {1790} (\bibinfo {year} {2013})}\BibitemShut {NoStop}%
\bibitem [{\citenamefont {Engelborghs}\ \emph {et~al.}(2002)\citenamefont
  {Engelborghs}, \citenamefont {Luzyanina},\ and\ \citenamefont
  {Roose}}]{ddebitool}%
  \BibitemOpen
  \bibfield  {author} {\bibinfo {author} {\bibfnamefont {K.}~\bibnamefont
  {Engelborghs}}, \bibinfo {author} {\bibfnamefont {T.}~\bibnamefont
  {Luzyanina}}, \ and\ \bibinfo {author} {\bibfnamefont {D.}~\bibnamefont
  {Roose}},\ }\href {\doibase 10.1145/513001.513002} {\bibfield  {journal}
  {\bibinfo  {journal} {{ACM} Transactions on Mathematical Software}\ }\textbf
  {\bibinfo {volume} {28}},\ \bibinfo {pages} {1} (\bibinfo {year}
  {2002})}\BibitemShut {NoStop}%
\bibitem [{\citenamefont {Izhikevich}(2010)}]{eugene}%
  \BibitemOpen
  \bibfield  {author} {\bibinfo {author} {\bibfnamefont {E.~M.}\ \bibnamefont
  {Izhikevich}},\ }\href
  {https://www.ebook.de/de/product/9425157/eugene_m_chairman_and_ceo_brain_corporation_izhikevich_dynamical_systems_in_neuroscience.html}
  {\emph {\bibinfo {title} {Dynamical Systems in Neuroscience}}}\ (\bibinfo
  {publisher} {MIT Press Ltd},\ \bibinfo {year} {2010})\BibitemShut {NoStop}%
\bibitem [{\citenamefont {Baer}\ \emph {et~al.}(1989)\citenamefont {Baer},
  \citenamefont {Erneux},\ and\ \citenamefont {Rinzel}}]{baer}%
  \BibitemOpen
  \bibfield  {author} {\bibinfo {author} {\bibfnamefont {S.~M.}\ \bibnamefont
  {Baer}}, \bibinfo {author} {\bibfnamefont {T.}~\bibnamefont {Erneux}}, \ and\
  \bibinfo {author} {\bibfnamefont {J.}~\bibnamefont {Rinzel}},\ }\href
  {\doibase 10.1137/0149003} {\bibfield  {journal} {\bibinfo  {journal} {{SIAM}
  Journal on Applied Mathematics}\ }\textbf {\bibinfo {volume} {49}},\ \bibinfo
  {pages} {55} (\bibinfo {year} {1989})}\BibitemShut {NoStop}%
\end{thebibliography}%

\end{document}